\documentclass[%
reprint,
superscriptaddress,
amsmath,amssymb,
aps,
prx,
longbibliography]{revtex4-1}

\usepackage{graphicx}% Include figure files
\usepackage{mathtools} %for over/under arrow
\usepackage{textcomp} %for \textdegree command
\usepackage[space]{grffile} %allows spaces in folder paths for graphics
\usepackage{dcolumn}% Align table columns on decimal point
\usepackage{bm}% bold math
\usepackage{textgreek} %for mu in text
\usepackage{titlesec}
\usepackage{xcolor}

\newcommand{\ket}[1]{\left| #1 \right>} % for Dirac kets
\newcommand{\titlename}{Telecom networking with a diamond quantum memory}

\begin{document}

\title{\titlename}

\author{Eric Bersin}
\thanks{These authors contributed equally to this work.}
\affiliation{Lincoln Laboratory, Massachusetts Institute of Technology, Lexington, MA 02421, USA}
\affiliation{Department of Electrical Engineering and Computer Science, Massachusetts Institute of Technology, Cambridge, MA 02139, USA}

\author{Madison Sutula}
\thanks{These authors contributed equally to this work.}
\affiliation{Department of Physics, Harvard University, Cambridge, MA 02138, USA}

\author{Yan Qi Huan}
\affiliation{Department of Physics, Harvard University, Cambridge, MA 02138, USA}

\author{Aziza Suleymanzade}
\affiliation{Department of Physics, Harvard University, Cambridge, MA 02138, USA}

\author{Daniel R. Assumpcao}
\affiliation{John A. Paulson School of Engineering and Applied Sciences, Harvard University, Cambridge MA 02138, USA}

\author{Yan-Cheng Wei}
\affiliation{Department of Physics, Harvard University, Cambridge, MA 02138, USA}

\author{Pieter-Jan Stas}
\affiliation{Department of Physics, Harvard University, Cambridge, MA 02138, USA}

\author{Can M. Knaut}
\affiliation{Department of Physics, Harvard University, Cambridge, MA 02138, USA}

\author{Erik N. Knall}
\affiliation{John A. Paulson School of Engineering and Applied Sciences, Harvard University, Cambridge MA 02138, USA}

\author{Carsten Langrock}
\affiliation{E. L. Ginzton Laboratory, Stanford University, Stanford CA 94305, USA}

\author{Neil Sinclair}
\affiliation{John A. Paulson School of Engineering and Applied Sciences, Harvard University, Cambridge MA 02138, USA}
%\affiliation{Division of PMA, and AQT, California Institute of Technology, Pasadena CA 91125, USA}

\author{Ryan Murphy}
\affiliation{Lincoln Laboratory, Massachusetts Institute of Technology, Lexington, MA 02421, USA}

\author{Ralf Riedinger}
\thanks{Current address: Institut fur Laserphysik und Zentrum f\"{u}r Optische Quantentechnologien, Universit\"{a}t Hamburg, 22761 Hamburg, Germany}
\affiliation{Department of Physics, Harvard University, Cambridge, MA 02138, USA}
%\affiliation{Institut f\"{u}r Laserphysik und Zentrum f\"{u}r Optische Quantentechnologien, Universit\"{a}t Hamburg, 22761 Hamburg, Germany}
%\affiliation{The Hamburg Centre for Ultrafast Imaging, 22761 Hamburg, Germany}

\author{Matthew Yeh}
\affiliation{John A. Paulson School of Engineering and Applied Sciences, Harvard University, Cambridge MA 02138, USA}

\author{C. J. Xin}
\affiliation{John A. Paulson School of Engineering and Applied Sciences, Harvard University, Cambridge MA 02138, USA}

\author{Saumil Bandyopadhyay}
\affiliation{Department of Electrical Engineering and Computer Science, Massachusetts Institute of Technology, Cambridge, MA 02139, USA}

\author{Denis D. Sukachev}
\affiliation{Department of Physics, Harvard University, Cambridge, MA 02138, USA}
\affiliation{AWS Center for Quantum Networking, Boston, MA 02135, USA}

\author{Bartholomeus Machielse}
\affiliation{Department of Physics, Harvard University, Cambridge, MA 02138, USA}
\affiliation{AWS Center for Quantum Networking, Boston, MA 02135, USA}

\author{David S. Levonian}
\affiliation{Department of Physics, Harvard University, Cambridge, MA 02138, USA}
\affiliation{AWS Center for Quantum Networking, Boston, MA 02135, USA}

\author{Mihir K. Bhaskar}
\affiliation{Department of Physics, Harvard University, Cambridge, MA 02138, USA}
\affiliation{AWS Center for Quantum Networking, Boston, MA 02135, USA}

\author{Scott Hamilton}
\affiliation{Lincoln Laboratory, Massachusetts Institute of Technology, Lexington, MA 02421, USA}

\author{Hongkun Park}
\affiliation{Department of Physics, Harvard University, Cambridge, MA 02138, USA}
\affiliation{Department of Chemistry and Chemical Biology, Harvard University, Cambridge, MA 02138, USA}

\author{Marko Lon\v{c}ar}
\affiliation{John A. Paulson School of Engineering and Applied Sciences, Harvard University, Cambridge MA 02138, USA}

\author{Martin M. Fejer}
\affiliation{E. L. Ginzton Laboratory, Stanford University, Stanford CA 94305, USA}

\author{P. Benjamin Dixon}
\affiliation{Lincoln Laboratory, Massachusetts Institute of Technology, Lexington, MA 02421, USA}

\author{Dirk R. Englund}
\affiliation{Department of Electrical Engineering and Computer Science, Massachusetts Institute of Technology, Cambridge, MA 02139, USA}

\author{Mikhail D. Lukin}
\affiliation{Department of Physics, Harvard University, Cambridge, MA 02138, USA}

\begin{abstract}
Practical quantum networks require interfacing quantum memories with existing channels and systems that operate in the telecom band.  
Here we demonstrate low-noise, bidirectional quantum frequency conversion that enables a solid-state quantum memory to directly interface with telecom-band systems. In particular, we demonstrate conversion of visible-band single photons emitted from a silicon-vacancy (SiV) center in diamond to the telecom O-band, maintaining low noise \mbox{($g^2(0)<0.1$)} and high indistinguishability \mbox{($V=89\pm8\%$).} 
%Furthermore, we perform conversion of telecom-band weak coherent state time-bin pulses to the visible band. 
We further demonstrate the utility of this system for quantum networking by converting telecom-band time-bin pulses, sent across a lossy and noisy 50~km deployed fiber link, to the visible band 
and mapping their 
%now attenuated single photon 
quantum states onto a diamond quantum memory with fidelity $\mathcal{F}=87\pm 2.5 \% $.
These results demonstrate the viability of SiV quantum memories integrated with telecom-band systems for scalable quantum networking applications.
\end{abstract}

\maketitle

Many promising applications of quantum information
systems 
%science 
%such as long-baseline interferometry~
\cite{Gottesman_2012, Khabiboulline_2019,
%}, secure communications~\cite{
Xu_2020,
%}, and distributed quantum computers~\cite{
Kimble_2008} require the ability to transmit quantum states over long distances. 
%Implementation of scalable techniques for long-distance quantum communication, in turn,  requires the careful integration of a wide range of quantum and classical components. 
Quantum memories with efficient, high-fidelity photon interfaces are a crucial enabling technology 
for long-distance quantum communication. They can serve as quantum network nodes, enabling single-photon generation~\cite{Lindner_2009}, realization of quantum repeaters~\cite{Briegel_1998,Munro_2010,Hermans_2022}, and deterministic quantum operations between photonic qubits~\cite{Schwartz_2016}.
%  mediated via matter qubits. 
%photon-photon interactions. 
While most promising quantum memories operate in the visible band, standard communication infrastructure 
%such as optical fibers and  integrated silicon photonic devices  
typically operates in the telecom band, designed for compatibility with existing communication channels such as low-loss deployed fibers.
% channels. 
%{\color{red} This incompatibility limits the incorporation of advanced capabilities offered by a wide range of quantum technologies into available and reliable telecom networks.} 
For these reasons, the realization of interfaces between visible-band quantum systems and telecom-band classical systems is  crucial for development of practical, networked quantum systems. 

Here, we realize a low-noise quantum frequency conversion (QFC)  system for interfacing a diamond silicon-vacancy (SiV) center quantum memory to the telecom O-band at 1350~nm. We demonstrate that this scheme can operate bidirectionally: first, we show that single photons generated by an SiV can be converted to 1350~nm while preserving their quantum properties. Next, we demonstrate that time-bin qubits generated at 1350~nm and subsequently converted to match the SiV's optical transition can be entangled with the SiV electron spin to enable high-fidelity heralded state transfer.

\begin{figure*}
\centering
\includegraphics{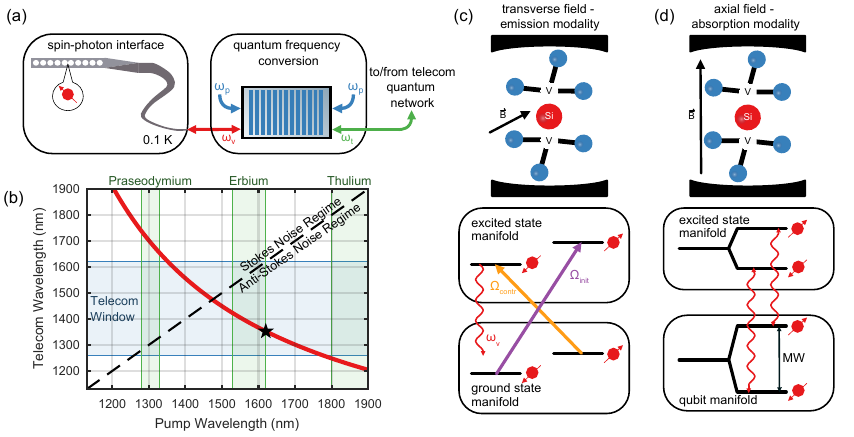}
\caption{(a)~Quantum networks based on  a spin-photon interface with a single atom-like defect center such as silicon-vacancy centers in diamond coupled to a nanophotonic cavity. Such memories often exhibit visible emission frequencies $\omega_v$, necessitating quantum frequency conversion (QFC) to a telecom frequency $\omega_t$ mediated by a pump at $\omega_p$.
(b) The parameter space for low-noise QFC using a single $\chi^{(2)}$-based conversion step. The red line indicates conversion schemes that can be performed for the SiV's 737~nm transition, the blue shaded region indicates the low-loss telecom window, and green shaded regions indicate what conversion processes can be easily pumped given common gain materials. The black star indicates our conversion scheme. (c)~SiV emission modality: an SiV in an overcoupled cavity and a transverse external magnetic field can be driven coherently with alternating initialization ($\Omega_{\text{init}}$) and control optical pulses ($\Omega_{\text{contr}}$) to emit single photons near 737~nm ($\omega_v$). (d)~SiV absorption modality: an SiV in a critically coupled cavity and an axial external magnetic field exhibits spin-conserving optical transitions that yield spin-dependent cavity reflectivity, enabling entanglement with an incoming photonic qubit at $\omega_v$.
}
\label{fig:overview}
\end{figure*}

In QFC, an input quantum field is overlapped with a strong pump field (frequency $\omega_p$) inside of a nonlinear material, shifting the frequency of the input while otherwise preserving its quantum properties~\cite{Kumar_1990}. This is often performed as depicted in Figure~\ref{fig:overview}(a), using difference(sum)-frequency generation in a $\chi^{(2)}$ material such as periodically poled lithium niobate (PPLN) to achieve frequency down(up)conversion between visible ($\omega_v$) and telecom ($\omega_t$) frequencies~\cite{Dreau_2018,Bock_2018,Ikuta_2018,Weber_2019}. While such nonlinear frequency conversion is routinely employed in classical applications for the production of coherent radiation 
at novel wavelengths or for detecting long-wavelength signals, in quantum applications, noise photons from spontaneous parametric downconversion (SPDC) and Raman scattering~\cite{Pelc_2010,Zaske_2011} generated by the strong pump field (typically 100~mW-class) lower the resulting fidelity of the transduced quantum state~\cite{Tchebotareva_2019}.
At frequencies below the pump frequency~$\omega_t<\omega_p$, Stokes Raman and SPDC processes result in a broadband noise plateau, %the noise levels plateau, with broadband noise from both Stokes Raman and SPDC processes, 
whereas at higher frequencies~$\omega_t>\omega_p$, the comparatively weaker anti-Stokes Raman signal typically tapers off 
%after around 30~THz above $\omega_p$
for separations $\Delta\omega_{tp}=\omega_t-\omega_p>30$~THz
\cite{Pelc_2011}.
This leads to two noise regimes --- one lower-noise, anti-Stokes regime where $\omega_p<\omega_t$, and one higher-noise, Stokes and SPDC-dominated regime where $\omega_p>\omega_t$.
Since $\omega_v$ is fixed by the choice of quantum memory, conservation of energy ${\omega_p = \omega_v-\omega_t}$ restricts the design of QFC schemes to only one degree of freedom. Moreover, the need for a strong pump laser introduces an additional engineering consideration due to availability of 
%of what pump lasers are readily available using 
common gain materials. The resulting trade-space (Figure~\ref{fig:overview}(b)) permits selection of a frequency conversion scheme based on the desired telecom wavelength, the noise requirements of the application, and the availability of pump laser amplifiers.

The negatively charged diamond SiV center is a promising candidate for the realization of quantum networks. Its electronic spin degree of freedom 
can store quantum information  over millisecond timescales at dilution refrigerator temperatures~\cite{Sukachev_2017}, and its optical transition exhibits nearly lifetime-limited spectral stability due to its inversion-symmetric electronic structure~\cite{Sipahigil_2014}. This stability permits integration into nanophotonic crystal cavities which provide an efficient interface for spin-photon state transduction.  
For example, as shown in Figure~\ref{fig:overview}(c),  an overcoupled photonic crystal cavity enhances emission  of the incorporated SiV into the desired optical mode. Under an applied transverse magnetic field, the SiV exhibits allowed transitions between spin states that can be driven to initialize the spin state via optical pumping, or to generate high quality, shaped single photons~\cite{Knall_2022}. This ``emission modality'' is a resource for quantum networking or quantum photonic processing applications which benefit from sources of high-quality single photons~\cite{Kok_2007}.
Alternatively, an SiV can operate in an ``absorption modality'' when incorporated into a critically coupled photonic crystal cavity and placed in an axial magnetic field as shown in Figure~\ref{fig:overview}(d)~\cite{Nguyen_2019}. This results in spin-dependent cavity reflectivity that can be used to generate entanglement with a reflected time-bin photonic qubit.
The SiV's 737~nm optical transition wavelength at which these protocols operate experiences strong loss in standard single-mode optical fiber ($>5$~dB/km), necessitating QFC to a low-loss telecom band. Examining the trade-space curve corresponding to the SiV in Fig.~\ref{fig:overview}(b), we identify a point (black star) where a long-wavelength pump at 1623~nm facilitates conversion to the edge of the O-band at 1350~nm (typical loss $\sim0.3$~dB/km in standard single-mode telecom fiber), permitting operation in the lower-noise anti-Stokes regime. This point maximizes $\Delta\omega_{tp}$ within the parameter space in order to minimize pump-induced noise. See Supplemental Figure~\ref{fig:the_plot} for a similar analysis using other emerging quantum memories, as well as with multi-pump conversion schemes~\cite{Hannegan_2021}.

\begin{figure*}
\centering
\includegraphics{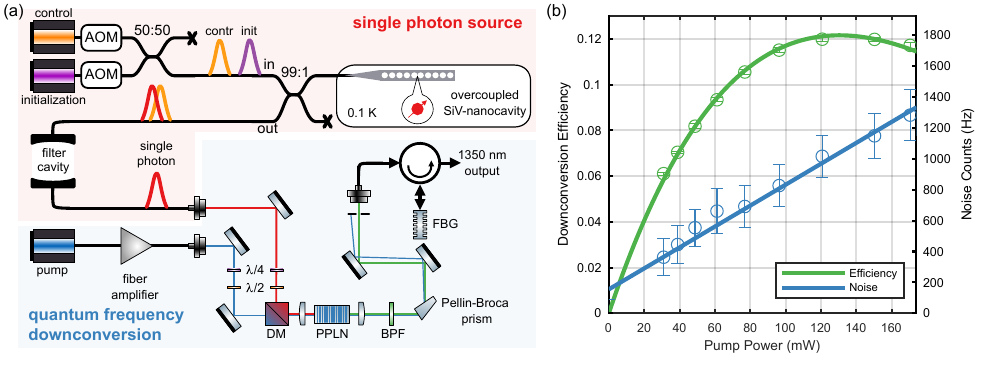}
\caption{
(a)~An SiV in emission modality is driven to produce single photons near 737~nm, which are efficiently collected by an overcoupled nanophotonic cavity then filtered from the driving fields with a free-space Fabry-P\'{e}rot cavity before being sent to a frequency downconversion setup. A pump laser at 1623~nm is amplified by a fiber amplifier (BKtel photonics) then coupled to free space. Half- and quarter-waveplates on both the pump and signal paths optimize coupling into the TE mode of the PPLN waveguide. The pump and signal are overlapped on a dichroic mirror (DM) and coupled into the PPLN waveguide where conversion occurs. The converted output is collected and filtered by a 25-nm-bandwidth bandpass filter (BPF), a Pellin-Broca prism, and ultimately a 50~GHz filter formed by a circulator and fiber Bragg grating (FBG). (b)~Total external conversion efficiency (green) measured using a classical 737~nm input, and pump-induced noise (blue) as a function of pump power inside the PPLN waveguide. The noise shown here is calculated for the output of the conversion setup, accounting for the efficiency and dark counts of our SNSPDs.}
\label{fig:downconversion}
\end{figure*}

Figure~\ref{fig:downconversion}(a) depicts our experimental setup for generating and converting single photons using a SiV center incorporated in an overcoupled photonic crystal cavity, following the approach used in Ref.~\cite{Knall_2022}. A tapered optical fiber is coupled to the device to deliver optical control pulses, as well as to efficiently collect single photons. We apply optical control pulses to produce a shaped single photon output with a temporal width of 30~ns at a repetition rate of 670~kHz. A Fabry-P\'{e}rot filter removes residual optical excitation light from the emitted photons, resulting in a final single photon generation efficiency of~$\eta=5.0\pm0.1\%$ within a 75~ns collection window. 

\begin{figure*}
\centering
\includegraphics{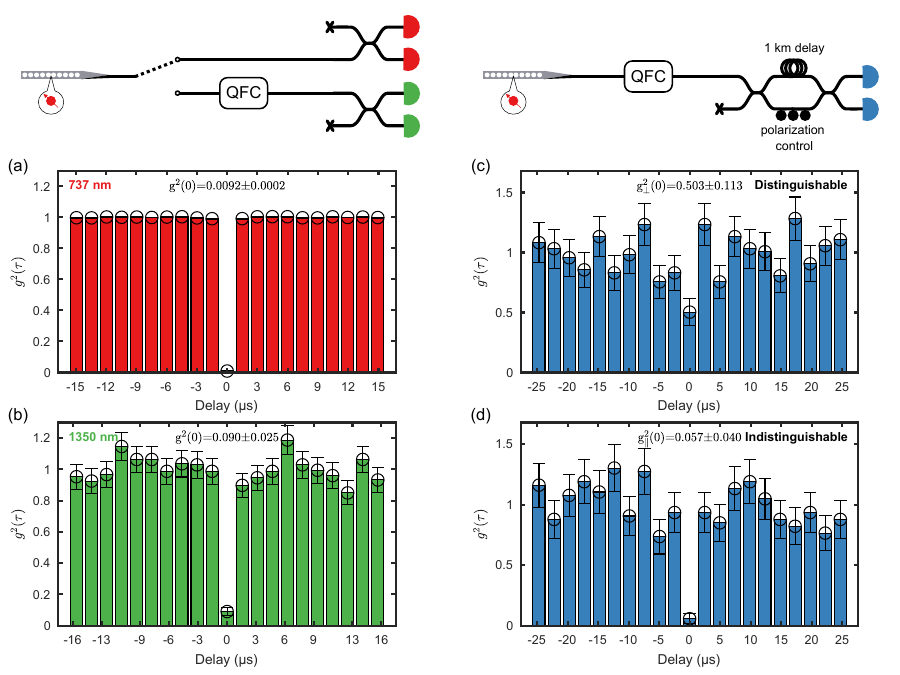}
\caption{Hanbury Brown-Twiss (HBT) interferometry of single photons emitted by an SiV (a)~before and (b)~after frequency downconversion, without correction for background noise or detector dark counts. In spite of the noise induced by the frequency converter due to our modest filtering, the $g^2(0)$ remains below $0.1$, indicating strong single photon quality. We further perform Hong-Ou-Mandel (HOM) interferometry between successive emissions of the SiV using a 1~km delay line, with a polarization controller in one arm of the delay line to compare between (c)~distinguishable and (d)~indistinguishable cases. The resultant visibility is measured to be $89\pm8$\%, in line with expectations from the background noise.}
\label{fig:single_photons}
\end{figure*}

Next, we convert these single photons to the telecom O-band using a custom-fabricated PPLN waveguide device (see Supplemental Material). Our input fields are a pump laser at 1623~nm amplified by a fiber amplifier
% then coupled into free-space. 
and a 737~nm signal, either from a laser for QFC efficiency characterization, or from the output of our SiV emission setup for single-photon conversion experiments. An aspheric lens couples both fields into separate waveguide modes of the PPLN, where an on-chip wavelength division multiplexer (WDM) combines the two modes. The PPLN waveguide output is collimated by a second aspheric lens, and residual pump light is removed by a 25~nm bandpass filter followed by a Pellin-Broca prism with a pinhole after 40~cm of propagation length. Finally, the converted light is sent through a fine filter setup comprising a circulator and a 50-GHz-bandwidth fiber Bragg grating to remove broadband anti-Stokes Raman noise.

In the strong-pump limit, the total external conversion efficiency is dependent on the pump power in the waveguide $P$
and the length of the interaction region $L$~\cite{Roussev_2004}:
\begin{equation}
\eta_{\text{ext}} = \eta_{\text{opt}}\eta_{\text{int}}\sin^2\left(L\sqrt{\eta_0P}\right),
\label{eq:conversion}
\end{equation}
where $\eta_{\text{opt}}$ is the combined efficiency of all optical components and filtering, $\eta_{\text{int}}$ is the maximum internal conversion efficiency of the PPLN, 
%$L$ is the length of the PPLN interaction region, 
and $\eta_0$ is the normalized internal efficiency, which is determined by the nonlinear material, mode overlap, etc. We measure $\eta_{\text{opt}}\approx19\%$, limited by coupling losses between the crystal waveguide mode and optical fiber (32\% total) and the fiber Bragg filter (59\%). By measuring the maximum depletion of the input 737~nm signal, we estimate an internal efficiency $\eta_{\text{int}}\approx65\%$. Sweeping the pump power (Fig.~\ref{fig:downconversion}(b)) and fitting the 1350~nm output power to Eq.~\ref{eq:conversion} shows a maximum conversion efficiency of $\eta_{\text{ext}}=12.2\pm0.1\%$, consistent with the losses described above. This conversion is maximized at an internal pump power of approximately 130~mW, which for our interaction length $L=35~$mm indicates a normalized internal efficiency $\eta_0\approx1.54\times10^4$~W$^{-1}$m$^{-2}$. These figures are comparable to existing demonstrations~\cite{Tchebotareva_2019,Krutyanskiy_2019,Leent_2020,Luo_2022}, and could be improved by further optimization of our coupling and device efficiency.

We probe the noise generated by the strong pump by measuring the setup output on superconducting nanowire single-photon detectors (SNSPDs) with zero 737~nm input power, plotted in Fig.~\ref{fig:downconversion}(b) after correction for the SNSPD detection efficiency.
The noise profile fits to a linear dependency of $6.8\pm0.9$~Hz/mW of noise photons generated by the pump, indicating anti-Stokes Raman scattering is the primary source. This noise reaches a count rate of $1.0\pm0.1$~kHz when the conversion efficiency is at its maximum.
Due to the broadband nature of this pump-induced noise, a useful metric is the noise spectral density at maximum conversion ${\rho=21\pm2}$~Hz/GHz, which compares favorably to other systems designed for diamond defect centers~\cite{Mann_2023} and indicates the ultimate fidelity limits for a QFC system given optimal filtering.

We characterize our QFC system's ability to preserve quantum states by measuring the second-order correlations of the single photon output. First, we benchmark our single photon source at its native visible wavelength, finding good single photon character of $g^2(0)=0.0092\pm0.0002$ without background or detector dark count subtraction (Fig.~\ref{fig:single_photons}(a)). Next, we convert these single photons to the telecom O-band, then send them through a 0.1-km-long fiber network (69\% efficient) between two buildings before detection on SNSPDs (29\% efficient). During the single-photon conversion experiments, we measure a mean QFC efficiency of $6\pm2\%$, attributing the additional loss and high variance compared to the classical case to polarization drift in the fibers between the SiV and our QFC setup. As a result, the signal-to-noise (SNR) ratio within our 75~ns gated time bins is  $14.5\pm0.1$.
For the converted photons, we measure $g^{(2)}(0)=0.090\pm0.025$, again without background or detector dark count subtraction (Fig.~\ref{fig:single_photons}(b)), indicating good preservation of quantum properties through QFC. 

\begin{figure*}
\centering
\includegraphics{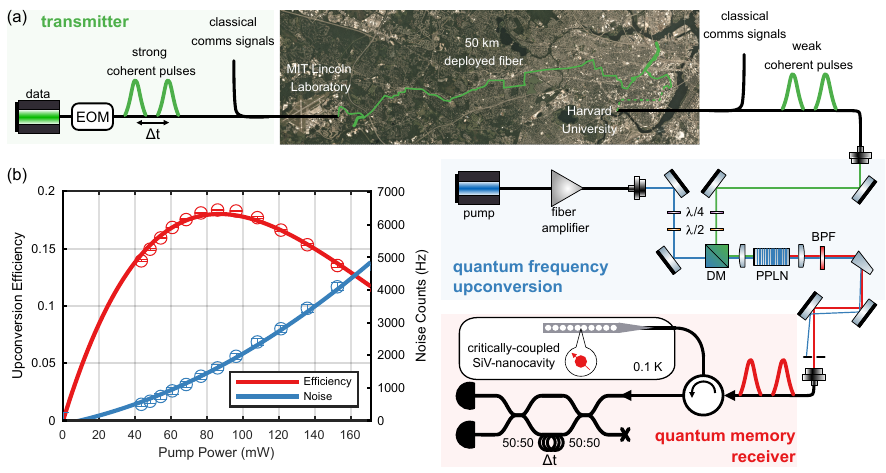}
\caption{
(a)~A 1350~nm data laser located at MIT Lincoln Laboratory is carved into time-bin pulses, multiplexed with signals for classical communications, then sent over a 50~km fiber link (green path: solid line exact route, dashed line approximate) which attenuates these pulses by 40.8~dB and adds environmental noise associated with deployed fiber. These data pulses are sent to a frequency upconversion setup designed similarly to the setup for downconversion~(Fig.~\ref{fig:downconversion}(a)), but without the narrow final FBG filter. The resultant visible time-bin qubits are sent through a circulator and reflected off an SiV in absorption modality. Reflected light is collected and measured on a time-delay interferometer, heralding state transfer. (b)~The relaxed filtering requirements permit an increased conversion efficiency of $18\pm0.2\%$, while maintaining a low noise count rate of $1.6\pm0.03$~kHz.}
\label{fig:upconversion}
\end{figure*}

The quantum properties of our converted photons can be further probed via Hong-Ou-Mandel interference between subsequent SiV emissions, using a 1~km delay line as shown in Figure~\ref{fig:single_photons}. A polarization controller in the short arm of the resultant interferometer permits us to adjust the relative polarizations between the two arms and compare between the distinguishable case, where the photons arriving from both arms at the final beamsplitter are cross-polarized (Fig.~\ref{fig:single_photons}(c)), and the indistinguishable case, where the photons arriving at the final beamsplitter are co-polarized (Fig.~\ref{fig:single_photons}(d)). We measure a HOM interference visibility $V=1-g^{(2)}_\parallel(0)/g^{(2)}_\perp(0) = 89\pm8\%$. 
Note that a comparison to the instrinsic indistinguishability pre-conversion is not possible due to the difficulty in retaining visible photons over the requisite km-scale delay line.

We next demonstrate that our scheme is compatible with absorption-based quantum memory protocols by performing frequency upconversion on a telecom-wavelength qubit. Figure~\ref{fig:upconversion}(a) depicts our experimental setup for generating telecom time-bin photonic qubits, upconverting them to match the SiV's visible transition, and generating spin-photon entanglement. 
For this experiment, we use an SiV in a critically coupled photonic crystal cavity~\cite{Stas_2022} and an axial magnetic field, resulting in spin-dependent cavity reflectivity that can be used to generate entanglement with a reflected time-bin photonic qubit.

We use a 1350~nm laser to generate coherent time-bin pulses with a full-width at half-maximum of 45~ns and separated by 144.5~ns.
%%%%%% Back-calculate how much light hitting the SiV from detection rate:
%got clicks in 2.1% of bins for ZZ, 2.5% for XX1, 1.4% for XX2 (same length so 2.0% total
%from SiV to detection is 0.143 WITHOUT circulator (from Science paper)
%circ out is 60%
% --> this gives 0.23 photons per qubit/per experiment hitting the SiV
These pulses are generated 
%at high intensity 
at MIT Lincoln Laboratory in Lexington, MA, 15.4~km away from the QFC receiver at Harvard University in Cambridge, MA. A 50~km deployed fiber link connects the two sites and attenuates the pulses by 40.8~dB such that they ultimately arrive at the SiV device as qubits with mean photon number $\langle\hat{n}\rangle\approx0.1$ per pulse. This channel also imparts noise associated with deployed fiber communications, including polarization drift, timing drift, and Doppler shifts. To coordinate between the two sites, we co-propagate a 1550~nm clock signal which synchronizes the experiment between the two sites, as well as a periodic 1350~nm reference signal to correct for polarization drifts over the deployed fiber and to serve as a reference for any drift in the photonic qubit's optical frequency. Details about the link as well as our classical communications protocol for synchronization and stabilization can be found in Ref.~\cite{Bersin_PRA_2023}.

\begin{figure}
\centering
\includegraphics{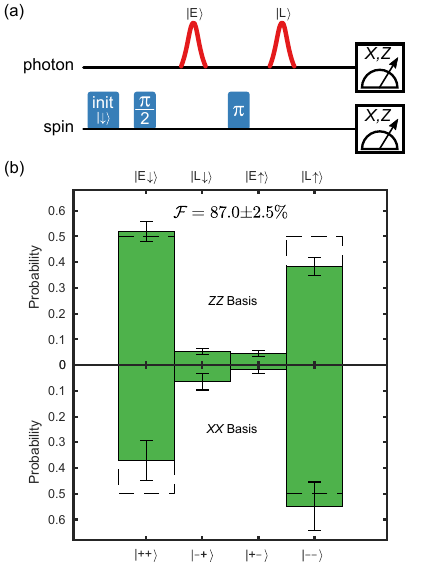}
\caption{(a) Pulse sequence for generating and verifying photonic state transfer. (b) Fidelity of the photon-to-spin mapping. Measurement of a transmitted photon in the $X$-basis heralds state transfer, whereupon the spin is read out in the $X$- or $Z$-basis, as appropriate for the photon state sent. 
}
\label{fig:state_transfer}
\end{figure}

We modify our QFC setup from Fig.~\ref{fig:downconversion}(a) to upconvert light from 1350~nm to 737~nm via sum-frequency generation using an identically-fabricated PPLN device.
Compared to the previous setup, we replace the lenses and filters to be wavelength-appropriate, and propagate the optical fields in reverse through the PPLN so that the on-chip WDM is at the output rather than the input.
Importantly, the increased spectral separation between the pump and converted wavelength for upconversion relaxes our filtering requirements, such that the converted light is only filtered by a 13-nm-bandwidth bandpass filter and a prism, without any fine filtering stage.
As shown in Figure~\ref{fig:upconversion}(b), we are thus able to attain an increased conversion efficiency of $18.0\pm0.2\%$ while maintaining a low noise count rate of $1.63\pm0.03$~kHz. We note that here, the noise profile fits well to a quadratic power dependency, suggesting these noise counts are dominated by upconversion of anti-Stokes photons rather than direct Raman noise~\cite{Pelc_2011}. The output of the conversion setup is sent through a fiber-optic circulator and delivered via a tapered fiber to the SiV-cavity, where light is transmitted or reflected depending on the SiV spin state. Reflected light is routed to a time-delay interferometer (TDI) 
that measures the photonic qubit using avalanche photodiodes.
Using this system and the pulse sequence shown in Figure~\ref{fig:state_transfer}(a), we generate the spin-photon entangled state ${\ket{\psi}=\left(\ket{E\downarrow}+\ket{L\uparrow}\right)/\sqrt{2}}$ conditioned on successful detection of a reflected photon.  Here, $\ket{E}$ and $\ket{L}$ denote the presence of a photon in an early or late time bin respectively.
We characterize this entanglement by measuring the photon and the spin in the joint $ZZ$- and $XX$-bases. Figure~\ref{fig:state_transfer}(b) shows the retrieved spin-photon correlations, limited primarily by residual reflectivity of the $\ket{\downarrow}$ state (see Supplemental Material). Following~\cite{Nguyen_2019_PRB}, we use these correlations to calculate an entanglement-mediated state transfer fidelity of $\mathcal{F}=87.0\pm2.5$\%. This result constitutes a heralded state transfer that reveals no information about the absorbed state, providing a key resource for  quantum repeaters and other quantum networking protocols.

Taken together, these experiments demonstrate bi-directional quantum state transfer between a diamond quantum memory and telecom photons facilitated by quantum frequency conversion. In particular, the SiV's 737~nm transition wavelength and our choice of conversion between the telecom O-band enabled the use of a long-wavelength pump, reducing the noise induced by the conversion process. Our heralded state transfer from a telecom photon to a diamond memory, incorporating classical telecommunication signals for timing synchronization and stabilization over deployed fiber, highlights the compatibility of the SiV platform with telecommunication infrastructure. This work can be extended and improved along several directions. 
%(Ian suggest: by an order of magnitude over ...)
The conversion efficiency was limited by coupling losses and non-unity internal conversion efficiency, and could be improved by further device optimization to realize up to $\eta_\text{ext}\approx70\%$~\cite{Yao_2020}. Our downconversion demonstration utilized a broad filter over $100\times$ the SiV's lifetime limited linewidth; using a narrower final filter thus offers a clear avenue for further SNR improvement. While our state transfer fidelity was limited by the finite cooperativity of the device used here (max $C\sim1.6$), other works have shown higher cooperativities ${C>100}$, enabling higher fidelities ${\mathcal{F}>94\%}$~\cite{Bhaskar_2020}.

The ability to transduce single photon quantum states between visible and telecom wavelengths with high fidelity has implications across quantum information science applications. Single telecom photons are an important resource for photonic quantum computing; replacing current SPDC-based sources with single telecom photons would improve processing efficiency and reduce errors from multi-photon events~\cite{Duan_2004,Knall_2022}. Furthermore, our system can be used for generating multiphoton entangled states when interfaced with a second cavity-coupled SiV~\cite{Bhaskar_2020}. When used as an input to a photonic processor, such cluster states can dramatically improve the efficiency and fidelity of photonic quantum computers~\cite{Lindner_2009}. When combined with access to nuclear memories~\cite{Stas_2022} and improved nanophotonic devices, these results offer a path towards more general deployed quantum networking.
 %using SiVs as quantum network nodes. 
 The high-bandwidth nature of our conversion system (see Supplemental Material) enables bridging of the frequency difference between individual SiVs due to the natural inhomogeneous distribution~\cite{Yu_2015,Sutula_2022,Stolk_2022}, and provides an avenue for taking advantage of the built-in spectral multiplexing of solid-state memories~\cite{Bersin_2019}. This approach also allows for interfacing SiVs with other quantum information platforms such as neutral atoms~\cite{Leent_2022}, trapped ions~\cite{Bock_2018,Krutyanskiy_2023}, or other color centers in diamond~\cite{Pompili_2021,Ruf_2021}, where performing frequency conversion to a common wavelength would facilitate photonic-mediated entanglement generation between these systems. This could enable quantum networks comprising  diverse  physical qubits, where different applications including computing, communication or sensing could be performed by physical systems best tailored to the task.
 %Denis comments:
 % 1 Comparison to other works?
 % 2 Applicability to other color centers/qubits
 % 3 Limitations, e.g. the consequences of g(2) = 0.1

\bibliography{references}

\begin{acknowledgments}
%Distribution Statement A. Approved for public release. Distribution is unlimited. 
This material is based upon work supported by the National Reconnaissance Office and the Under Secretary of Defense for Research and Engineering under Air Force Contract No. FA8702-15-D-0001. Any opinions, findings, conclusions or recommendations expressed in this material are those of the authors and do not necessarily reflect the views of the National Reconnaissance Office or the Under Secretary of Defense for Research and Engineering. \textcopyright~2023 Massachusetts Institute of Technology. Delivered to the U.S. Government with Unlimited Rights, as defined in DFARS Part 252.227-7013 or 7014 (Feb 2014). Notwithstanding any copyright notice, U.S. Government rights in this work are defined by DFARS 252.227-7013 or DFARS 252.227-7014 as detailed above. Use of this work other than as specifically authorized by the U.S. Government may violate any copyrights that exist in this work.

We thank Catherine Lee for helpful discussions.
This work was supported by the National Science Foundation (NSF, Grant No. PHY-2012023), NSF EFRI ACQUIRE (Grant No. 5710004174), NSF QuIC-TAQS (Grant No. OMA-2137723), Center for Ultracold Atoms (Grant No. PHY-1734011), Department of Energy (DoE, Grant No. DESC0020115), AFOSR MURI (Grants No. FA9550171002 and No. FA95501610323), and Center for Quantum Networks (Grant No. EEC-1941583). Devices were fabricated at the Harvard Center for Nanoscale Systems, NSF award no. 1541959. E.B. and M.S. acknowledge funding from a NASA Space Technology Research Fellowship. Y.Q.H. acknowledges support from the Agency for Science, Technology and Research (A*STAR) National Science Scholarship. D.A., E.N.K., and B.M. acknowledge support by the NSF Graduate Research Fellowship under Grant No. DGE1745303. M.Y. acknowledges funding from the Department of Defence (DoD) through the National Defense Science and Engineering Graduate (NDSEG) Fellowship Program. We thank Planet.com for allowing us to take custom satellite imagery. 

\end{acknowledgments}

%\section*{Author Contributions}

%%%%%%% Begin Supplemental materials %%%%%%%

\clearpage
\widetext
\begin{center}
\textbf{\large Supplemental Material for ``\titlename''}
\end{center}
%%%%%%% Prefix a "S" to all equations, figures, tables and reset the counter %%%%%%%
\setcounter{equation}{0}
\setcounter{section}{0}
\setcounter{figure}{0}
\setcounter{table}{0}
\setcounter{page}{1}
\makeatletter
\renewcommand{\theequation}{S\arabic{equation}}
\renewcommand{\thefigure}{S\arabic{figure}}
\renewcommand{\thesection}{S\arabic{section}}
\renewcommand{\bibnumfmt}[1]{[S#1]}
\renewcommand{\citenumfont}[1]{#1}
%%%%%%%%%% Prefix a "S" to all equations, figures, tables and reset the counter %%%%%%%%%%

\section{Frequency Conversion Tradespace}

\begin{figure*}[h]
\centering
\includegraphics{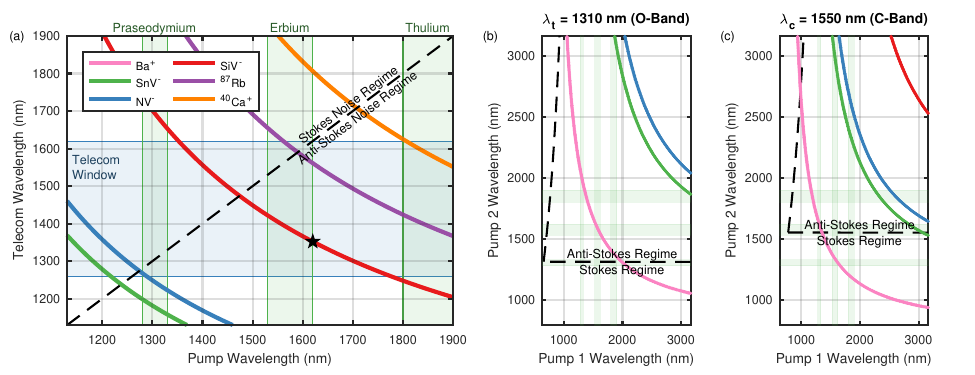}
\caption{\textbf{Quantum Frequency Conversion Tradespace.} (a) Conservation of energy restricts $\chi^{(2)}$-based quantum frequency conversion to a single degree of freedom due to the fixed native optical wavelength of a given quantum memory, such as a given species of trapped ions (${\lambda_{\text{Ba}^+}=493}$~nm~\cite{Hannegan_2021}, ${\lambda_{^{40}\text{Ca}^+}=854}$~nm~\cite{Bock_2018}), diamond defect centers (${\lambda_{\text{SnV}^-}=619}$~nm~\cite{Trusheim_2020}, ${\lambda_{\text{NV}^-}=637}$~nm~\cite{Tchebotareva_2019}, ${\lambda_{\text{SiV}^-}=737}$~nm), or neutral atoms (${\lambda_{^{87}\text{Rb}}=780}$~nm~\cite{Leent_2022}). Combined with the desire to use a pump wavelength where gain media are available to produce high pump powers (X-axis), to convert between a telecom wavelength in the low-loss telecom window (Y-axis), and to operate in the low-noise anti-Stokes regime, the resultant parameter space poses different challenges for various quantum memories. Using two conversion pumps to convert to 1310~nm~(b) or 1550~nm~(c) can open up additional options. Here, the X-axis is the wavelength of Pump 1, the Y-axis is the wavelength of Pump 2, the green bands are once again the regions with well-developed gain materials, and the black dashed line separates the two noise regimes.}
\label{fig:the_plot}
\end{figure*}

As discussed in the main text, quantum frequency conversion poses a tradespace where loss, noise, and availability of laser gain material must be balanced. Figure~\ref{fig:overview}(b) depicts this for the SiV; Figure~\ref{fig:the_plot}(a) repeats this figure, adding curves for other emerging quantum memories. This tradespace assumes a single $\chi^{(2)}$-based frequency conversion step; an alternative approach is to use either a higher order $\chi^{(3)}$-based process where $\omega_t=\omega_v-\omega_{p1}-\omega_{p2}$, or multiple $\chi^{(2)}$-based conversion stages, in either case using two pumps at frequencies $\omega_{p1}$ and $\omega_{p2}$~\cite{Esfandyarpour_2018,Hannegan_2021}. This is attractive for systems that need to span large frequency differences, as it can permit each individual step to operate in the low-noise anti-Stokes regime. Here, while most quantum systems can operate in the anti-Stokes regime, a greater challenge is the dearth of common gain materials for the requisite pump laser wavelengths. Figures~\ref{fig:the_plot}(b)--(c) plot the parameter space for these ``two-step'' conversions. Due to the larger solution space for this approach, we fix the final converted wavelength to either 1310~nm (Fig.~\ref{fig:the_plot}(b)) or 1550~nm (Fig.~\ref{fig:the_plot}(c)) corresponding to the telecom O- and C-bands respectively. We note that these telecom bands are $\sim100$~nm wide each, giving some flexibility in the vicinity of these curves. Importantly, this figure omits consideration of the challenges that arise for large separations between $\lambda_{v}$ and $\lambda_{p}$, such as engineering strong overlap between waveguide modes when one wavelength is much larger than the other, which can lead to undesired multimode behavior of the shorter $\lambda_v$ (though this can also be used intentionally~\cite{Lu_2021}). 

\section{Periodically Poled Lithium Niobate (PPLN) Device}

\begin{figure*}
\centering
\includegraphics{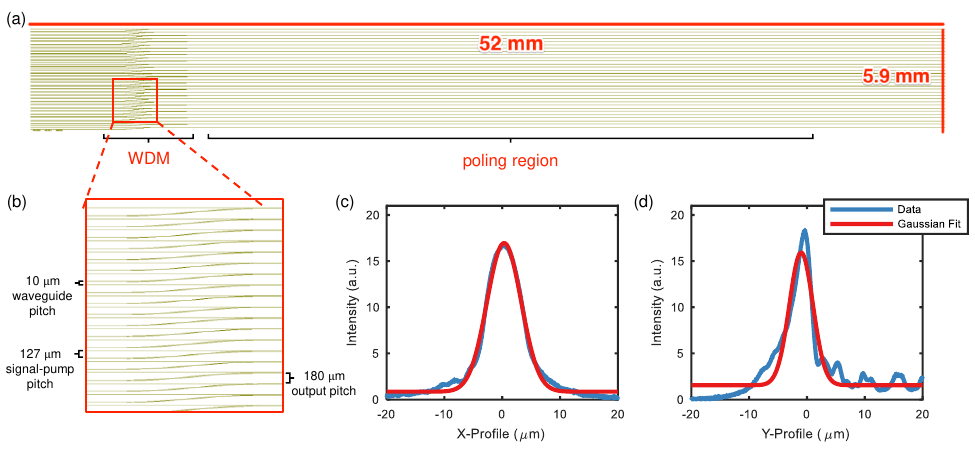}
\caption[PPLN device layout]{\textbf{PPLN device layout} \textbf{a)}, Schematic of the PPLN device showing the 52~mm-long RPE waveguide with a wavelength division multiplexer (WDM) at the input followed by a 35-mm-long poling region to facilitate conversion. \textbf{b)}, The WDM comprises waveguides designed for signal (737~nm) and pump (1623~nm) respectively, fabricated with a 127~\textmu m pitch. These are brought together with a directional coupler which serves as a WDM, resulting in a final pitch of 180~\textmu m. Imaging the output waveguide mode at 1350~nm using a 100$\times$ lens, the horizontal profile of the mode fits well to a symmetric Gaussian (\textbf{c}), while the vertical profile exhibits an asymmetry, which may be an interference artifact (\textbf{d}).}
\label{fig:chip}
\end{figure*}

We fabricate a device starting from bulk lithium niobate to facilitate QFC, using periodic poling to achieve quasi-phasematching (QPM) and incorporating waveguides to realize efficient spatial overlap of the participating optical modes. As shown in the schematic in Figure~\ref{fig:chip}(a), the length of the reverse-proton-exchange (RPE) periodically poled lithium niobate (PPLN) waveguide chip, containing 32 individual QFC devices, was 52~mm. Each QFC device comprised an on-chip directional coupler (Fig.~\ref{fig:chip}(b)) with a waveguide-to-waveguide separation of 10~\textmu m to efficiently combine the 1623~nm pump and the 737~nm signal, and a 35-mm-long QPM section with a poling period of 14.5 um for Type-0 difference-frequency generation between the 737-nm signal and the 1623-nm pump. The waveguide width in the coupler section was 5~\textmu m, while it was 8~\textmu m in the QPM section, optimized for noncritical operation~\cite{Bortz_1994}. 
%A typical transfer function for one of these devices is shown in Fig. YY. 

To maximize the conversion and coupling efficiency, we designed the two 127-\textmu m-separated input ports to each device (pump and signal) to only support the respective lowest-order waveguide modes and have mode-field diameters close to those encountered in single-mode fiber. Imaging the output mode of these waveguide devices at 1350~nm, we find that the waveguide mode is fairly Gaussian in the horizontal direction (x-direction), with a 1/$e^2$ mode field diameter (MFD) of 11.8~\textmu m (Fig.~\ref{fig:chip}(c)). However, we observe an asymmetry in the vertical (y-direction) extent of the mode with an 8.2~\textmu m MFD (Fig.~\ref{fig:chip}(d)), decreasing coupling efficiency between this mode and the Gaussian mode of optical fiber. An antireflection coating was also applied (Forreal Spectrum) to each end facet to improve input coupling efficiency. While we operated these device using free-space coupling, low-loss fiber coupling has previously been demonstrated and is expected to introduce a 0.5~dB coupling loss for both pump and signal, which could improve the external conversion efficiency up to $\eta_\text{ext}\approx70\%$~\cite{Yao_2020}.

\begin{figure*}
\centering
\includegraphics{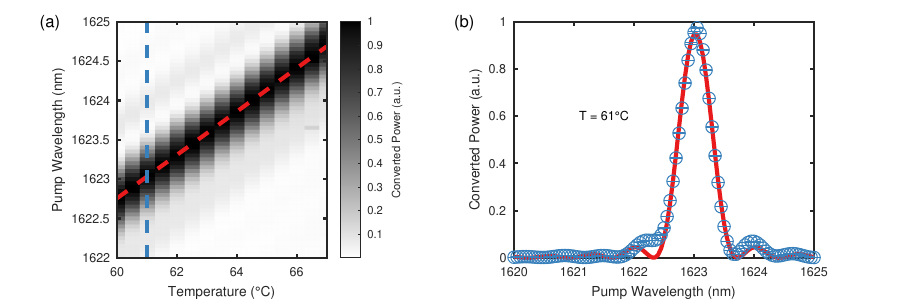}
\caption[PPLN transfer function]{\textbf{PPLN transfer function} \textbf{a)}, Normalized conversion efficiency as a function of chip temperature $T$ and pump wavelength, showing tunability of the phase-matching condition by 0.28~nm/$^\circ$C (red dashed line). \textbf{b)} A cross-section of the transfer function at $T=61^\circ$C (blue dashed line in \textbf{a}). Fitting the expected sinc function to the data (data in blue, fit in red) shows a phase-matching bandwidth of around 0.8~nm.}
\label{fig:temptune}
\end{figure*}

We characterized the transfer function of one waveguide on the device by measuring the power of converted light as a function of chip temperature and pump power, shown in Figure~\ref{fig:temptune}(a). As we did not realign the chip in between temperature setpoints, the results have been normalized for each temperature setting to account for differences in coupling efficiency. We find a 0.28~nm/$^\circ$C shift in the phase-matched pump wavelength as a function of chip temperature (red dashed line). A cross-section of this transfer function is shown in Figure~\ref{fig:temptune}(b), taken at the temperature ($T=61^\circ$C) where our desired conversion process was optimally phase-matched. A fit to the expected sinc function shows a slight asymmetry in the transfer function, indicating 
%slight imperfections from ideal poling~\cite{Bortz_1994}; 
minor geometrical inhomogeneities of the waveguide along its length that locally modify the dispersion; 
however, the function largely matches theoretical expectations, indicating the device is well-suited for efficient frequency conversion. The experiments described in the main text were performed on a waveguide which had an optimal phase-matching temperature of 55.5$^\circ$C, which we achieved by operating the chip inside of a temperature-controlled housing (HC Photonics).

%\subsection{Single photon emission}
%Single photon generation was done using the three level protocol detailed in Ref.~\cite{Knall_2022}, which also contains details about device fabrication techniques and cavity parameters. For single photon generation, the SiV’s electron is first initialized in the $m_s=\ket{\uparrow}$ state. The SiV is then pumped on the electron spin flipping transition, leading to an adiabatic transfer of electron population to the $\ket{\downarrow}$ state and the generation of a photon via the cavity enhanced ($C=10$) spin conserving transition.

\section{State Absorption Experiment Details}

\subsection{Deployed Fiber Network}

For the state absorption experiment, our transmitter (Alice) is stationed at MIT Lincoln Laboratory in Lexington, MA. Our memory-enabled receiver (Bob) is stationed at Harvard campus in Cambridge, MA; this node is connected by 50~km of fiber to Alice (40.8~dB at 1350~nm). A third network node (Charlie) used only as a passive pass-through in this work but available for future demonstrations is located at MIT campus in Cambridge, MA. A 43~km strand of single-mode telecom fiber (24.1~dB at 1350~nm) connects Alice to Charlie, and 7~km of fiber connects Charlie to Bob (19.7~dB at 1350~nm), forming an all-to-all 3-node network. We note that these losses are primarily due to splices in this commercial fiber link rather than the expected inherent exponential propagation loss. For extended details on this link, see Ref.~\cite{Bersin_PRA_2023}.

\subsection{Transmitter}
\begin{figure*}
\centering
\includegraphics{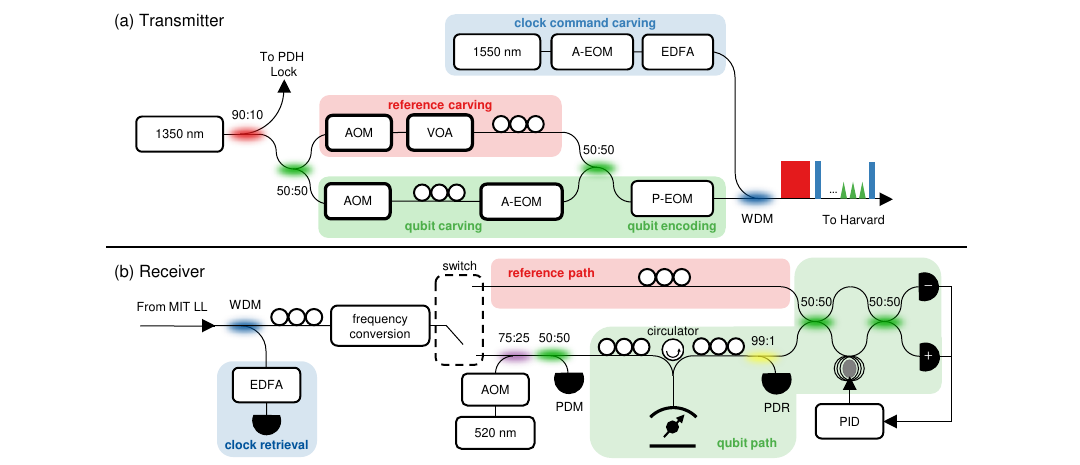}
\caption[Full transmitter optics schematic.]{\textbf{Full Tx/Rx optics schematic.} a) A 1350~nm laser, locked to an ultra low expansion reference cavity via a Pound-Drever-Hall (PDH) scheme, is split into two paths --- one for carving strong reference pulses using an acousto-optic modulator (AOM) and MEMS variable optical attenuator (VOA), and one for carving weak qubit pulses using an AOM and an amplitude electro-optic modulator (A-EOM). These signals are combined, then encoded with data using a phase electro-optic modulator (P-EOM) and multiplexed with the clock command signals, which are generated via amplitude modulation and amplification (EDFA) of a 1550~nm laser. b) Incoming signals from Alice are first demultiplexed to separate clock pulses from other signals, the former of which are detected on a fast photodiode (PDC) for sequence and timing synchronization. Qubits and reference signals are upconverted, and a MEMS switch routes reference light directly to the TDI, while qubits are routed toward the memory. A green 520~nm signal is multiplexed in to allow qubit reinitialization, and two photodetectors coupled by a 50:50 fiber beamsplitter enable us to monitor input power (PDM) and reflected power (PDR). After bouncing off the memory, facilitated by a fiber-optic circulator (42\% total efficiency), photonic qubits are detected on the TDI.}
\label{fig:full_optics}
\end{figure*}

\begin{figure*}
\centering
\includegraphics{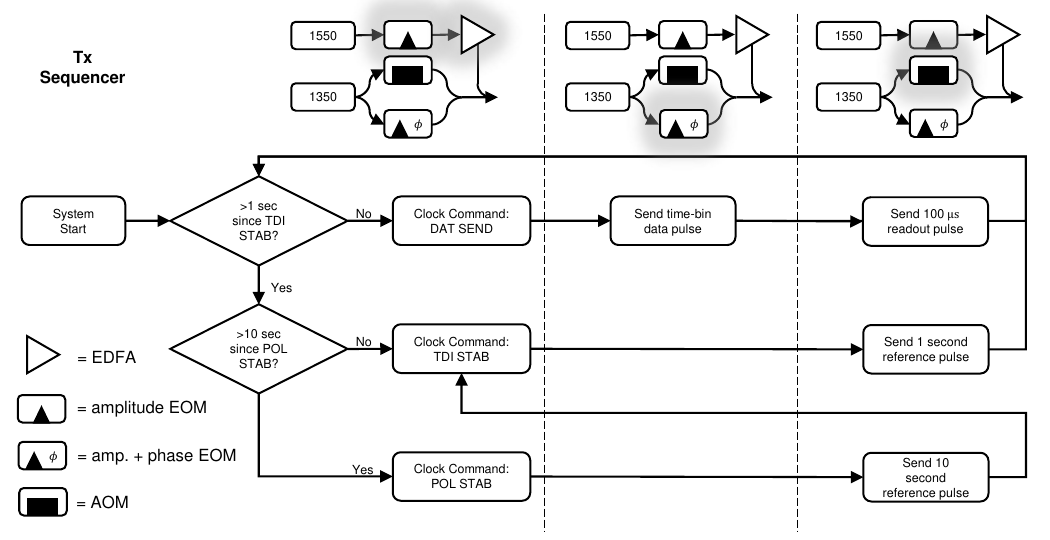}
\caption[Transmitter experimental sequence.]{\textbf{Transmitter experimental sequence.} Logical flowchart for the AWG sequencer used to control Alice's transmitter. Inset diagrams show a simplified version of the transmitter hardware, with glows indicating components which are being actively controlled during given portions of the sequence.}
\label{fig:sequence_tx}
\end{figure*}

The overall optical setup used for generating the quantum and classical signals sent from Alice to Bob is shown in Figure~\ref{fig:full_optics}(a). Details about the classical signals for performing timing synchronization, polarization drift correction, and frequency transfer across the link can be found in Ref.~\cite{Bersin_PRA_2023}. Our photonic qubits are generated by attenuated coherent laser pulses. A 1350~nm ECDL laser (Toptica) is locked to an ultra low expansion Fabry-P\'{e}rot cavity (Stable Laser Systems) to ensure narrow linewidth. Light from this laser is split out, passed through an accousto-optic modulator (AOM, Brimrose) for power and frequency control, then passed through an amplitude electro-optic modulator (EOM, EOSPACE, Inc.) which carves pulses to form time bin qubits. Each pulse is a Lorentzian with full width at half maximum $\Gamma=45$~ns, and the time bins are spaced 144.5~ns apart. After pulse carving, a phase EOM encodes phases on each time bin, achieved by applying a step-like voltage pattern to the EOM. The resultant output is combined with our synchronization and reference signals before being sent across the network. All of the opto-electronic devices for the transmitter are controlled using a single HD Arbitrary Waveform Generator (HD-AWG, Zurich Instruments). The full experimental sequence for this transmitter AWG is shown in Figure~\ref{fig:sequence_tx}.

\subsection{Receiver: Optics and Sequencing}
\label{sec:qkd_rx}
\begin{figure*}
\centering
\includegraphics[width=\linewidth]{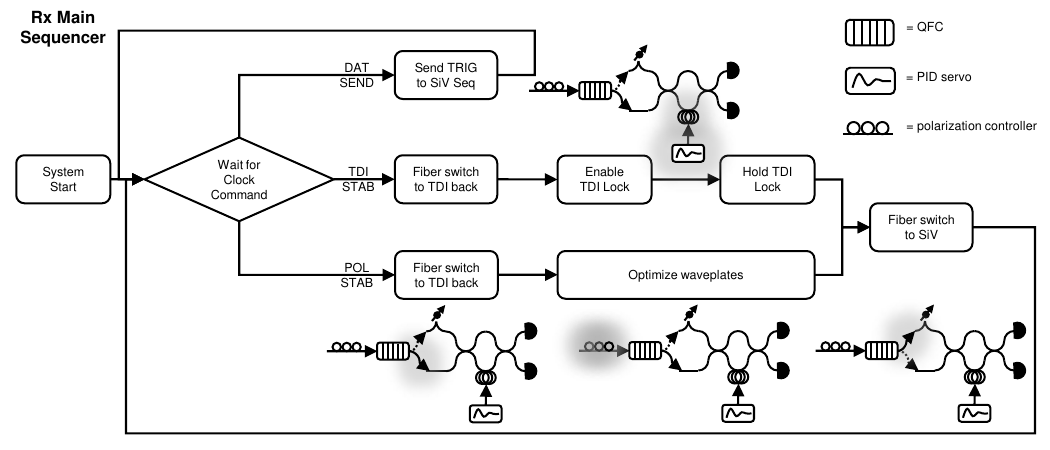}
\caption{\textbf{Receiver experimental sequence - Main Sequencer.} Logical flowchart for the AWG sequencer used to control the overall sequence of Bob's receiver, with decision trees based on the clock command signals received from Alice. Inset diagrams show a simplified version of the transmitter hardware, with glows indicating components which are being actively controlled during given portions of the sequence. This sequencer also triggers the SiV sequencer (Fig.~\ref{fig:sequence_rx_siv}), which runs a subroutine for SiV microwave and optical control.}
\label{fig:sequence_rx_main}
\end{figure*}

The overall optical setup used at the receiver for processing incoming clock, reference, and data signals is shown in Figure~\ref{fig:full_optics}(b). The receiver makes use of two AWG sequencers: one ``Main Sequencer'' that processes classical information arriving over the fiber network and dictates the progress of the overall experimental sequence (Fig.~\ref{fig:sequence_rx_main}; for detailed operation, see Ref.~\cite{Bersin_PRA_2023}), and one ``SiV Sequencer'' that is triggered by the Main Sequencer and is responsible for SiV control (see below). As stated above, Alice's transmitter coordinates between periods of data transmission and stabilization by sending clock commands to Bob, where they are picked off, amplified, detected, digitized, and processed by Bob's Main Sequencer according to the flow chart shown in Fig.~\ref{fig:sequence_rx_main}. During data periods, a MEMS switch (Agiltron) routes incoming qubits along the ``qubit path'' towards the memory. During stabilization periods, the MEMS switch bypasses the memory and routes the incoming high intensity reference signal directly to the TDI via the ``reference path'' to avoid fridge heating and excessive radiation on the SiV.

\subsection{Receiver: Diamond network node}
\label{sec:diamond_node}
\subsubsection{Optics}
Details about the SiV-cavity system, including fabrication techniques, cavity parameters, and spin properties can be found in Ref.~\cite{Stas_2022}, which employed the same device and SiV defect as the experiments discussed here, though on a separate cool-down of the system, such that the precise values of the cavity QED parameters, coupling efficiencies, etc. may slightly differ. As shown in Fig.~\ref{fig:full_optics}(b), the optical path for our qubits after the MEMS switch comprises a 75:25 splitter for incorporating 520~nm repump light (with qubits following the 75\% path), a 50:50 splitter for measuring incident and reflected powers, various polarization controllers, a fiber-optic circulator (Ascentta), the SiV-cavity memory, and a TDI for qubit measurement. We measure 9\% transmission efficiency through the fiber-optic components (MEMS switch, splitters, and circulator), and estimate the efficiency of the SiV-cavity reflectivity (including losses from the adiabatic tapered fiber coupling) to be 15\%. Combined with our frequency conversion efficiency of 19\%, this yields an overall efficiency to detect photons arriving at the receiver node of 2.6$\times10^{-3}$.

\subsubsection{Electronics}
\begin{figure*}
\centering
\includegraphics[width=\linewidth]{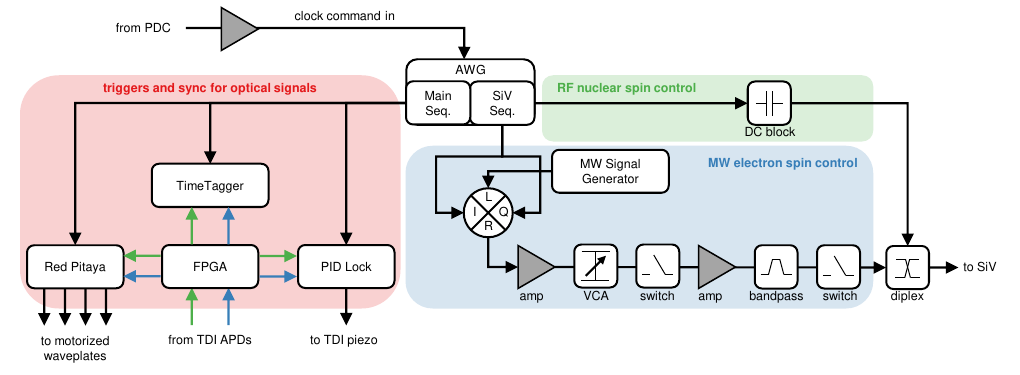}
\caption[Electronics for SiV receiver.]{\textbf{Electronics for SiV receiver.} Clock commands from Alice detected on the PDC (Fig.~\ref{fig:full_optics}) trigger the AWG, which controls the remaining electronics. The AWG runs two sequencers: the first is a ``Main Sequencer'' (Main Seq.) that coordinates the overall experiment via digital lines for counting and triggering signals related to the optical signals, including photon detection, timing synchronization, and responding to classical reference light (red region). A second ``SiV Sequencer'' (SiV Seq.) controls the SiV via microwave (MW) lines for electron spin control (blue region) and radio-frequency (RF) lines for nuclear spin control (green region).}
\label{fig:rx_electronics}
\end{figure*}

Figure~\ref{fig:rx_electronics} shows the electronics used to control the SiV spin and all other necessary components in the receiver. The components of this figure related to SiV spin control and detection of reflected photons has been adapted from Ref.~\cite{Stas_2022}, which contains additional information about the parameters used for these electronics. Electron spin control is performed by the microwave (MW) electronics shown in the blue region of Fig.~\ref{fig:rx_electronics}, where an radio-frequency (RF) output of the AWG is mixed with a local oscillator, then amplified, filter, and switched before being sent to the dilution refrigerator, where gold striplines facilitate spin driving. Nuclear spin control is achieved by direct driving from the AWG's RF output (green region of Fig.~\ref{fig:rx_electronics}).

The red region of Fig.~\ref{fig:rx_electronics} shows circuitry related to the optical signals of the experiment. Detected TTL signals from our avalanche photodiodes (APDs) are sent to a custom-built FPGA which, which simply serves as a buffer to split the signal to three other systems: the AWG for state initialization logic, a TimeTagger (Swabian Instruments) for photonic qubit heralding and SiV state readout, and a custom-built PID Lock box that generates an error signal from the difference in APD counts. This PID box also produces a buffered voltage proportional to the total count rate, which is sent to a Red Pitaya; this enables the Red Pitaya to control the automated waveplates on our conversion setup to maximize system throughput, thus correcting for polarization drifts over the deployed fiber. Finally, the activity of these systems is controlled by digital lines from the AWG which trigger these various protocols and serve as a timing reference for the TimeTagger. Not shown here, digital signals from the AWG are also used to control the MEMS switch and AOM for our 520~nm repump laser as shown in Fig.~\ref{fig:full_optics}(b).

\subsubsection{Spin-Cavity System}

\begin{figure*}
\centering
\includegraphics{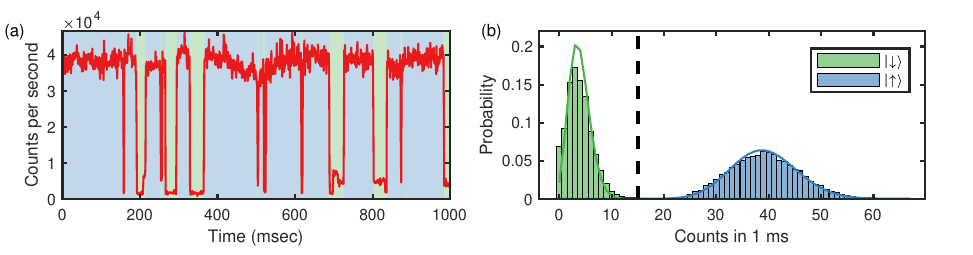}
\caption[Spin readout.]{\textbf{SiV-Cavity reflectivity contrast.} a) Free-running count rate of weak light reflected off of the SiV-nanocavity system, showing distinct quantum jumps between the SiV spin states that correspond to periods of high (blue) and low (green) reflectivity. b) Example probability distributions of counts received during 1~ms-long readout bins, where each portion of the bimodal distribution (colored green and blue) has been independently normalized. The bimodal distribution in observed counts corresponds to the two SiV spin states, whose strong coupling to the cavity results in spin-dependent reflection. Fitting Poisson probability distributions to each portion of the distribution (green and blue lines), we find that a threshold of 15 counts (black dashed line) permits single-shot state readout with an error of $6\times10^{-6}$.}
\label{fig:contrast}
\end{figure*}

Both our spin readout and our spin-photon gate rely on high reflectivity contrast between the bright and dark SiV spin states. Figure~\ref{fig:contrast}(a) shows a free-running time trace of reflected counts off of the SiV-nanocavity system under weak illumination. Distinct jumps between periods of high and low reflectivity are observed, corresponding to quantum jumps in the SiV spin state. This behavior enables us to perform high-fidelity, single-shot, non-demolition spin readout under pulsed illumination of the system. Figure~\ref{fig:contrast}(b) also shows an example histogram of the counts retrieved during such 1~ms readout bins. For this dataset, the SiV was initialized into $\ket{+}$ before readout. Here, the data have been separated out into two categories (green and blue) based on the clear bimodal behavior, and each of the two resultant distributions normalized and fit with a Poisson probability distribution. Processing the data in this way, where the state of the SiV is purposefully placed into a superposition before readout, permits us to isolate our determination of the readout fidelity from any infidelities in state initialization. For the data shown here, the retreived mean counts of $\lambda_\downarrow=3.86$ and $\lambda_\uparrow=39.1$ indicate that by thresholding at 15 counts, single-shot readout can in principle be performed with an error probability of $6\times10^{-6}$. Note that despite the high readout fidelity, the infidelity of our spin-photon gate will still be limited by the contrast $\lambda_\downarrow/\lambda_\uparrow=9.9\%$.

\begin{figure*}
\centering
\includegraphics[width=\linewidth]{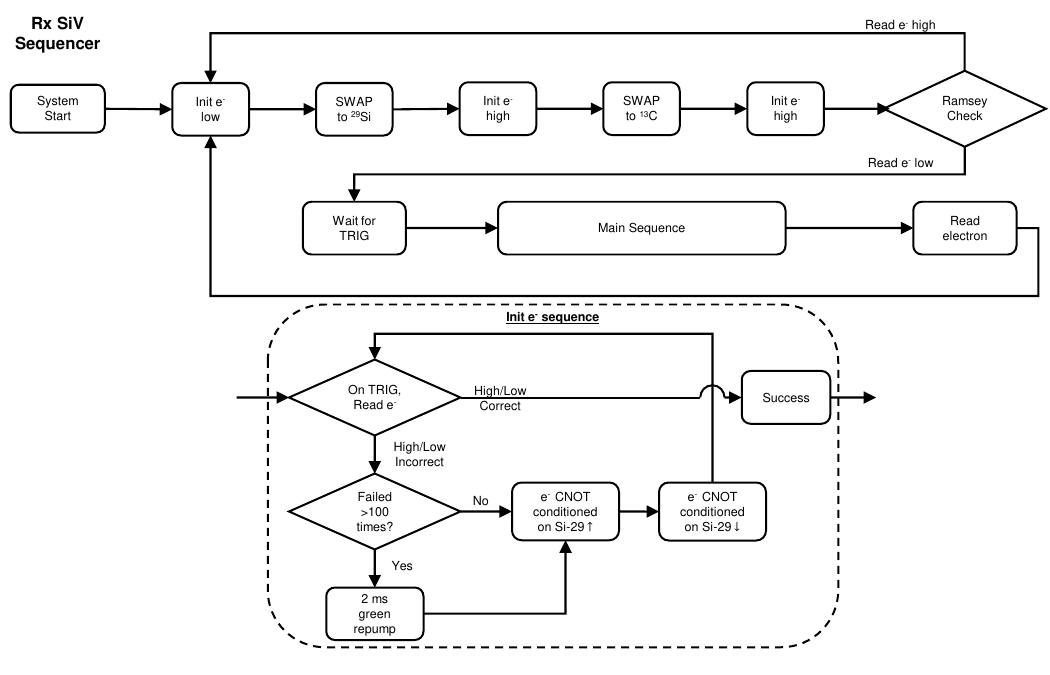}
\caption[Receiver experimental sequence - SiV Sequencer.]{\textbf{Receiver experimental sequence - SiV Sequencer.} Logical flowchart for the AWG sequencer used specifically for microwave and optical control of Bob's SiV. On system start, the sequencer will initialized the SiV spin, following the sequence in the dashed box for either ``low'' (corresponding to the low reflectivity $m_s=\ket{\downarrow}$ state) or ``high'' (corresponding to the high reflectivity $m_s=\ket{\uparrow}$ state). It then waits for a trigger from the Main Sequencer (Fig.~\ref{fig:sequence_rx_main}), at which point it runs a Main Sequence for spin-photon interactions --- typically the sequence shown in Fig.~\ref{fig:state_transfer}(a).}
\label{fig:sequence_rx_siv}
\end{figure*}

Due to spectral diffusion of the SiV transition, this contrast can drift over time, diminishing the fidelity of both our readout and our spin-photon gate. Before every string of photonic qubits, we thus perform an SiV initialization sequence that verifies good contrast $>10$ between the bright and dark states, only moving forward once this check has cleared.  The full sequence for SiV spin initialization is shown in Figure~\ref{fig:sequence_rx_siv}. For this check, we read the state of the spin using a 100~\textmu s pulse of readout light; if we are attempting to initializing into $\ket{\downarrow}$ ($\ket{\uparrow}$), this check only passes if we measure fewer (more) than 2 (20) photons. Otherwise, we attempt to reinitialize the spin, following the sequence depicted in Fig.~\ref{fig:sequence_rx_siv}. Furthermore, in order to ensure high contrast for the incoming qubits themselves, all spin readout for this initialization sequence is performed using upconverted reference light from Alice, rather than using a laser local to Bob. As such, all readout --- both during the main experimental sequence and during initialization --- is timed based on the incoming clock commands from Alice.

\subsubsection{Spin control fidelity}

\begin{figure*}
\centering
\includegraphics{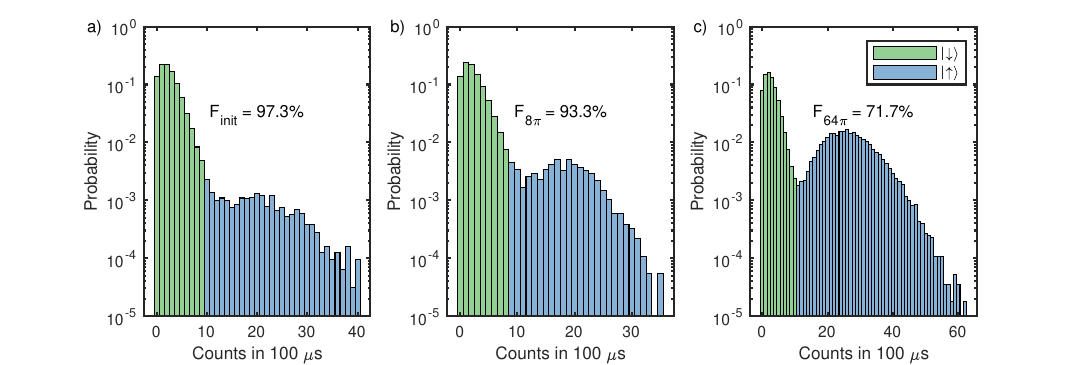}
\caption[Spin control fidelity]{\textbf{Spin control fidelity.} Histograms of photon counts during 100~\textmu s readout bins after various gate sequences. For each plot, the readout threshold is determined by fitting two Poisson distributions to the data as described above, such that counts below (above) threshold, shaded green (blue), are considered a measurement of $\ket{\downarrow}$ ($\ket{\uparrow}$). Each sequence was designed to end with the spin in $\ket{\downarrow}$ in the case of no errors, such that each sequence's fidelity is simply the sum of the green-shaded area. a) Readout histogram after initializing the spin into $\ket{\downarrow}$. b) Readout histogram after initializing the spin into $\ket{\downarrow}$, then applying 8~\textpi-pulses (an $XY8$ decoupling sequence) spaced by 144.5~ns. c) Readout histogram after initializing the spin into $\ket{\downarrow}$, then applying 64~\textpi-pulses (8-$XY8$ sequences) spaced by 144.5~ns.}
\label{fig:fidelities}
\end{figure*}

We characterize our control of the SiV spin by performing a variety of gate sequences that, when error-free, should conclude with the SiV in $\ket{\downarrow}$. First, we perform the initialization sequence described above, using 100~\textmu s pulses for initialization and readout. We find that this sequence initializes the spin with $97.3\%$ fidelity (Fig.~\ref{fig:fidelities}(a)). Next, we check the fidelity of our microwave gates by performing a modified version of the XYN sequence used for the main experiment, where the initial and concluding $\pi/2$ gates are removed to reduce sensitivity to external noise. The results of this check are shown in Fig.~\ref{fig:fidelities}(b) for XY8 and Fig.~\ref{fig:fidelities}(c) for XY64. The degradation in fidelity here indicates a gate fidelity of 99.5\% per \textpi-pulse,
%calculated this by doing:
%   (0.933/0.973)^(1/8) = 99.48% and
%   (0.717/0.973)^(1/64) = 99.52%
which provides an upper limit on the fidelity we can expect to achieve for longer pulse sequences.

\end{document}